\documentclass[a4paper,11pt]{article}
\usepackage[T1]{fontenc}
\usepackage[utf8]{inputenc}
\usepackage{lmodern}
\usepackage{graphicx}
\usepackage[hidelinks]{hyperref}
\usepackage{csquotes}
\usepackage{array}

\newcommand{\articleTitle}{The EU AI Act is a good start but falls short}
\newcommand{\articleSubtitle}{}
\newcommand{\articleAuthors}{
    Chalisa Veesommai Sillberg\(^{a,}\)\authorFootnoteOrcid{chalisa.sillberg@tuni.fi}{0000-0001-8400-7469},
    Jose Siqueira De Cerqueira\(^{a,}\)\authorFootnoteOrcid{jose.siqueiradecerqueirag@tuni.fi}{0000-0002-8143-1042},
    Pekka Sillberg\(^{a,}\)\authorFootnoteOrcid{pekka.sillberg@tuni.fi}{0000-0003-2573-4775},
    Kai-Kristian Kemell\(^{a,}\)\authorFootnoteOrcid{kai-kristian.kemell@tuni.fi}{0000-0002-0225-4560},
    Pekka Abrahamsson\(^{a,}\)\authorFootnoteOrcid{pekka.abrahamsson@tuni.fi}{0000-0002-4360-2226}
}
\newcommand{\articleAffiliation}{
    \(^{a}\) Tampere University, Faculty of Information Technology and Communication Sciences, Pori, Finland \\
}
\newcommand{\articleDate}{August 26, 2024}

\newcommand{\articleNotes}{Notes: 
    \begin{itemize}
        \item This is the author's version of the work
        \item This is a pre-print version of an article accepted to the 15\(^{th}\) International Conference on Software Business (ICSOB 2024)
        \item The final version accepted to publishing the Springer Lecture Notes in Business Information Processing (LNBIP).
    \end{itemize}
}

\newcommand*{\authorFootnote}[1]{\footnote{email: \url{#1}}}
\newcommand*{\authorFootnoteOrcid}[2]{\footnote{email: \url{#1}; ORCID iD: \href{https://orcid.org/#2}{[#2]}}}
\usepackage{xspace}
\newcommand*{\orcidID}[1]{\href{https://orcid.org/#1}{[#1]\xspace}}

\usepackage{tabularx}
\usepackage{booktabs}
\usepackage{multirow}

\begin{document}



\pagenumbering{roman}\setcounter{page}{1} 
\begin{titlepage}

\begin{figure}[ht]
    \centering
    \includegraphics[height=2em]{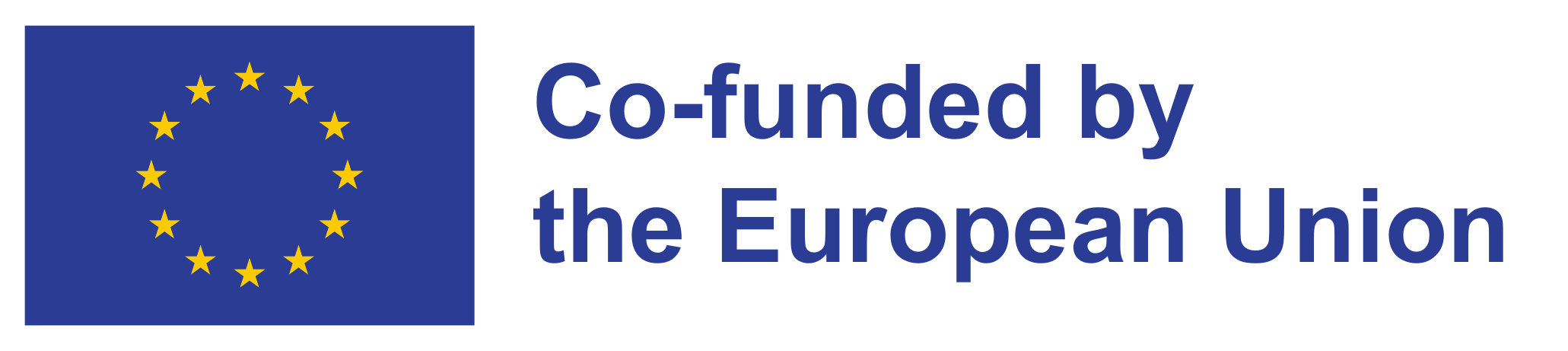}
        \hspace{1em}
    \includegraphics[height=2em]{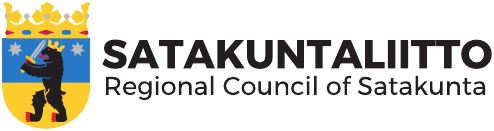}
        \hspace{1em}
    \includegraphics[height=2em]{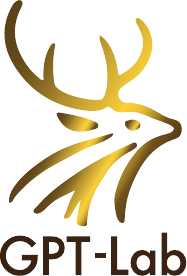}
        \hspace{1em}
    \includegraphics[height=2em]{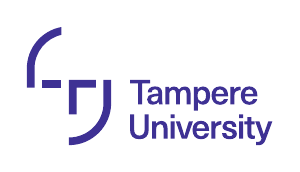}
\end{figure}

    \centering
        \vspace{2em}

        \noindent\Large
		\begingroup
			\textbf{\articleTitle}
		\endgroup
  
        \vspace*{0.5em}
        \noindent\large
		\begingroup
			\textbf{\articleSubtitle}
		\endgroup

\vfill

 \noindent\large
            \articleAuthors

\vspace*{0.5em}
 \noindent\normalsize
		\articleAffiliation

        \vspace*{0.5em}
            \articleDate

\vfill

\articleNotes

\vfill

\end{titlepage}

\clearpage

\clearpage
\pagenumbering{arabic}\setcounter{page}{1} 



%
\title{The EU AI Act is a good start but falls short}

\author{
    Chalisa Veesommai Sillberg\(^a\)\orcidID{0000-0001-8400-7469} \and 
    Jose Siqueira De Cerqueira\(^a\)\orcidID{0000-0002-8143-1042} \and
    Pekka Sillberg\(^a\)\orcidID{0000-0003-2573-4775} \and 
    Kai-Kristian Kemell\(^a\)\orcidID{0000-0002-0225-4560} \and 
    Pekka Abrahamsson\(^a\)\orcidID{0000-0002-4360-2226} \and
    \\
    \(^a\) Tampere University, Faculty of Information Technology and \\ Communication Sciences,
    Pori, Finland \\
    \texttt{chalisa.sillberg@tuni.fi} 
}
\date{\vspace{-5ex}} 

\maketitle


\section*{Abstract}
The EU AI Act was created to ensure ethical and safe Artificial Intelligence development and deployment across the EU. This study aims to identify key challenges and strategies for helping enterprises focus on resources effectively. To achieve this aim, we conducted a Multivocal Literature Review (MLR) to explore the sentiments of both the industry and the academia.
From 130 articles, 56 met the criteria. Our key findings are three-fold. First, liability. Second, discrimination. Third, tool adequacy.
Additionally, some negative sentiments were expressed by industry and academia regarding regulatory interpretations, specific requirements, and transparency issues. Next, our findings are three essential themes for enterprises. First, risk-based regulatory compliance. Second, ethical frameworks and principles in technology development. Third, policies and systems for regulatory risk management.
These results identify the key challenges and strategies and provide less commonly discussed themes, enabling enterprises to align with the requirements and minimize their distance from the EU market.

\section{Introduction}
The European Commission launchs the first European Union's (EU) regulatory framework for Artificial Intelligence (AI), namely the EU AI Act (EU AIA), in April 2021 \cite{marinkovic2023new}. The EU AIA aims to regulate AI while preserving innovation and upholding fundamental rights, a challenging balance to achieve. It sets out four specific objectives: 1) to ensure that AI systems in the EU market comply with safety laws and respect fundamental rights and Union values; 2) to promote a legal framework that encourages investment and innovation in AI; 3) to improve the effective enforcement of laws related to fundamental rights and safety in AI systems; and 4) to facilitate the development of a single market for safe, legal and trustworthy AI systems, preventing market fragmentation \cite{heymann2023operating}. 

To achieve this goal, the EU AIA proposes a proportionate regulatory approach with the minimum requirements necessary in terms of the risks and potential problems associated with AI. The EU AIA outlines prohibited AI systems, explains the requirements for high-risk AI systems, and provides guidelines to increase transparency while fostering development. Performing a risk classification is the first step toward knowing the risk category and complying with regulatory requirements. The EU AIA categorizes AI systems into three risk levels: 1) unacceptable risk, which includes AI functionalities such as subliminal manipulation, exploitative techniques, biometric categorization, social scoring, real-time remote biometric identification, emotional state assessment, predictive policing, and facial image scraping, all of which are banned in the EU; 2) high risk, which includes AI systems such as those used to assess consumer creditworthiness; and 3) limited and minimal risk, which includes AI chatbots and similar functionalities \cite{eu2021a}. In the context of the EU AIA, enterprises, i.e., providers, deployers, importers, distributors, and product manufacturers, are faced with detailed documentation, strategies, and processes related to compliance. These may include various practical, operational, and strategic issues. Therefore, the process and framework should be in the form of practical terms. According to current law, the Secretary of Government Operations is required to create a coordinated plan that looks into the viability and challenges of creating standards and technology that state agencies can use to identify the provenance of digital content by analyzing the effects of the rise in deepfakes, among other things \cite{ScottW2024}.

Since the emergence of the EU AIA, several academic articles have been published to share good practics, such as the challenges posed by AI-driven trading in the EU's financial markets \cite{azzutti2022ai}, the effect on mobility within the EU [GL34: see Data availability], a framework for collaborative governance \cite{outeda2024eu}, the articulation of ethical charters, legal tools, and technical documentation in Machine Learning (ML)[WL24: see Data availability]. Furthermore, valuable insights may be available not in traditional academic databases but in the media, such as the EU AIA's recommendations for business [GL39], a proposed framework with current developments meant for businesses, a comprehensive regulatory framework for businesses \cite{Sophie2023,Frida2024}, and implications and strategies for UK businesses\cite{Christopher2023}, to name but a few. Thus, this work recognises that valuable knowledge and insights can be found in various types of publications and media, not just academic articles \cite{garousi2019guidelines}. Accordingly, we aim to aid companies in better understanding EU AIA by conducting a secondary research study using a Multivocal Literature Review (MLR) and Natural Language Processing (NLP) to investigate the EU AIA practices.

In this study, we aim to gather a more comprehensive and multifaceted understanding of the EU AIA by considering multiple perspectives, including those from practitioners, industry experts, and other stakeholders. For this reason two research questions (RQs) were formulated for conducting this study: RQ1) \emph{What are the key challenges perceived by both industry and academia in complying with the EU AIA}, and RQ2) \emph{What strategies and processes are enterprises developing to implement the EU AIA}, through MLR and NLP approaches. MLR is a comprehensive literature review approach and systematic process that incorporates various sources of information, namely "Grey Literature" (GL). MLR has become notable in various fields because it is an approach able to 1) integrate both academic and GL, 2) provide a comprehensive overview of a topic with diverse perspectives, and 3) raise the strictness and applicability in the literature review of that field for proving the significance of practical and knowledge \cite{garousi2019guidelines,kamei2021grey}. 

This study is structured similarly to the MLR guidelines \cite{garousi2019guidelines}, in terms of 1) planning and surveying the MLR guidelines, 2) conducting the review and data analysis, and 3) presenting the findings or results. The rest of the paper is organized as follows. The content in Section \ref{methodology} describes 1) how to search, select, and stop data sources with criteria, 2) how to control the quality of the data used, and 3) how to analyze the key challenges and identify/underline the strategies. The content in Section \ref{result} deals with the key challenges and the strategy themes that will benefit companies in complying with the EU AIA. Section \ref{discussion} describes and provides a practical framework and its limitations. Finally Section \ref{conclusionandfuturework} concludes the study and suggests future research work.

\section{Methodology}
\label{methodology}
In this study, a systematic MLR and NLP approach to EU AIA in business was applied in three parts. In the first part, a process for conducting the review and gathering the significant data was established. It also looked outside of academic forums, according to the search process criteria, and made data preparation and formation for determining the relevant data without overlaps and duplicate data, in accordance with the selection criteria and performance of the data selection. In the second part, quality was assessed for determining the satisfactory and free of bias nature of the data source. Lastly, data was analyzed to find the key challenges, and underline the themes of the strategies and processes being developed to implement the EU AIA.

\subsection{Search strategy and literature review protocol}
There are several specific data sources related to the EU AIA available in scientific databases. In practice, the vast array of data sources with valuable insights also exists outside the realm of scientific publications, often obtained by practitioners and not discussed or published in the scientific literature. Therefore, an MLR approach to systematic review is beneficial as it incorporates GL.

\textbf{Search data source}. Four digital databases, i.e., ScienceDirect, IEEE Xplore, ACM Digital Library, and Google Scholar, were considered for searching white literature (WL). However, IEEE Xplore provided insufficient results relevant to EU AIA. Google search engine was utilized for the GL search.

\textbf{Search strings}. Keyword search is crucial for efficiently locating relevant information by targeting specific terms. Effective keyword searches help filter out irrelevant data, allowing researchers to focus on pertinent literature. To establish relevant keywords, a systematic approach was used:
 	\emph{Basic Keywords:} Initially, direct keywords related to the research topic were identified, including "EU," "AI," "Act," and "business".
	\emph{Phrase Search:} These basic keywords were then combined into a phrase to derive relevant keyword searches, resulting in the phrase \texttt{"EU AI Act" AND "business"}.
 	\emph{Web Search Engine:} The phrase search was employed in a general search engine, specifically Google, which is a conventional tool for web search and GL review studies across many fields.

\textbf{Pre-selection}. To narrow down the vast amount of available data sources, a pre-selection process was applied to enhance the likelihood of finding relevant, high-quality sources. This process aimed to cover all aspects of the research topic comprehensively. Inclusion data (ID) and exclusion data (ED) criteria were defined for selecting data sources and are depicted in Figure~\ref{fig:Process1}.

\textbf{Stop search}. To ensure a thorough and efficient search, stopping criteria for the GL search were applied in this study, referred to as control stopping criteria. Two criteria were used: effort bounded and bound limitation with a condition. The effort-bounded criterion limited the search to the first 100 search engine hits, with the option to extend if exclusion data (ED) were found among these hits. Additionally, the ratio of relevant results to total search results was monitored. If this ratio fell below 0.5, indicating the bound limitation condition, the search process was terminated \cite{butijn2020blockchains}.

\subsection{Quality assessment for the literature review approach}

The quality assessment criteria were reviewed and set to ensure the selected data sources were relevant, unbiased, and of high quality. Some GL quality assessment processes overlap with inclusion/exclusion criteria, add exclusions, and are integrated with study assessments. Therefore, we used the advanced criteria from Garousi for this purpose, as outlined in Table~\ref{table:QualityAssessmentCriteria}. Then, the dataset of GL data sources was transformed into a CSV format for database creation and quality assessment analysis.

\begin{table}[ht]
    \caption{Quality assessment criteria of GL for EU AI Act in Business \cite{garousi2019guidelines}}
    \centering
    \scriptsize
    \begin{tabularx}{\textwidth}{ p{2.5cm} X p{0.850cm} }
   \toprule
  Criteria 								& Exclusion Questions (EQ) 						& Satisfy\\
  \midrule

\multirow{4}{=}{1. Authority}	& 1. Is the publishing organization reputable? & \multirow{4}{*}{3/4} \\
 			& 2. Is the individual author associated with a reputable organization?\\
 			& 3. Has the author published other work in the field? \\
			& 4. Does the author have expertise in the area?
        \\ \midrule
\multirow{6}{=}{2. Methodology} &   1. Does the source have a clearly started aim?	& \multirow{6}{*}{4/6} \\
 				& 2. Does the source have a stated methodology? \\
 				& 3. Is the source supported by authoritative, contemporary references? \\
				& 4. Are any limits clearly started? \\
				& 5. Does the work cover a specific question? \\
				& 6. Does the work refer to a particular population or case? \\ \midrule
\multirow{4}{=}{3. Objectivity} 	& 1. Does the work seem to be balanced in presentation?	& \multirow{4}{*}{4/4} \\
 				& 2. Is the statement in the sources as objective as possible, or is the statement a subjective opinion? \\
 			    & 3. Is there vested interest, or are the conclusions free of bias? \\
				& 4. Are the conclusions supported by the data? \\ \midrule
4. Date 		& 1. Does the item have a clearly stated date?	 & 1/1\\ \midrule
5. Position w.r.t. related source & 1. Have key related GL or formal sources been linked to / discussed? & 1/1\\ \midrule
\multirow{2}{=}{6. Novelty}     & 1. Does it enrich or add something unique to the research? & \multirow{2}{=}{2/2} \\
 				& 2. Does it strengthen or refute a current position? \\  \midrule
7. Impact		& 1. Normalize following impact metrics into a single aggregated value: citations, backlinks, social media shares, comments,
view counts	  & 1/1\\  \midrule
\multirow{3}{=}{8. Outlet type} 	& $ 1^{st} $ tier GL (m=1) is high outlet control and high credibility
        & \multirow{3}{*}{0-1} \\
 				& $ 2^{nd} $ tier GL (m=0.5) is moderate outlet control and credibility \\
 				& $ 3^{rd} $ tier GL (m=0) is low outlet control and low credibility \\
    \bottomrule
    \end{tabularx}
    \label {table:QualityAssessmentCriteria}
    \end{table}


\subsection{Data analysis approach}
The aim of this approach was to gain a comprehensive understanding of the key challenges and strategies perceived by both industry and academia in complying with the EU AIA. We utilized an approach similar to the one utilized by Bourdin\cite{bourdin2024nlp}. This approach was conducted using Altair AI Studio software (version 2024.0.1), which combines both qualitative and quantitative, providing an analysis of the research question. 

\textbf{Data extraction}. A summary and key takeaway context of each data source were extracted to obtain a broad overview of all key points and emphasize the most critical point or lesson gained from the content.

\textbf{Data analysis}. The NLP algorithm is part of AI and has been created for numerous purposes. It aims to achieve human language understanding, focusing on enabling computers to analyze and interpret text data \cite{pilowsky2024natural,loor2024systematic}. There are two analysis approaches in this part: the first process analyzes the key challenges by implementing text processing, Term Frequency-Inverse Document Frequency (TF-IDF), and sentiment analysis in the summary context, and the second process identifies and underlines the theme of strategies and processes by applying text processing, TF-IDF, and the Latent Dirichlet Allocation (LDA) model in the key takeaway context. The data analysis process of this study is shown in Figure~\ref{fig:Process1}, and is described as follows:

\begin{figure}[ht]
	\centering
	\includegraphics[width=1.0\textwidth]{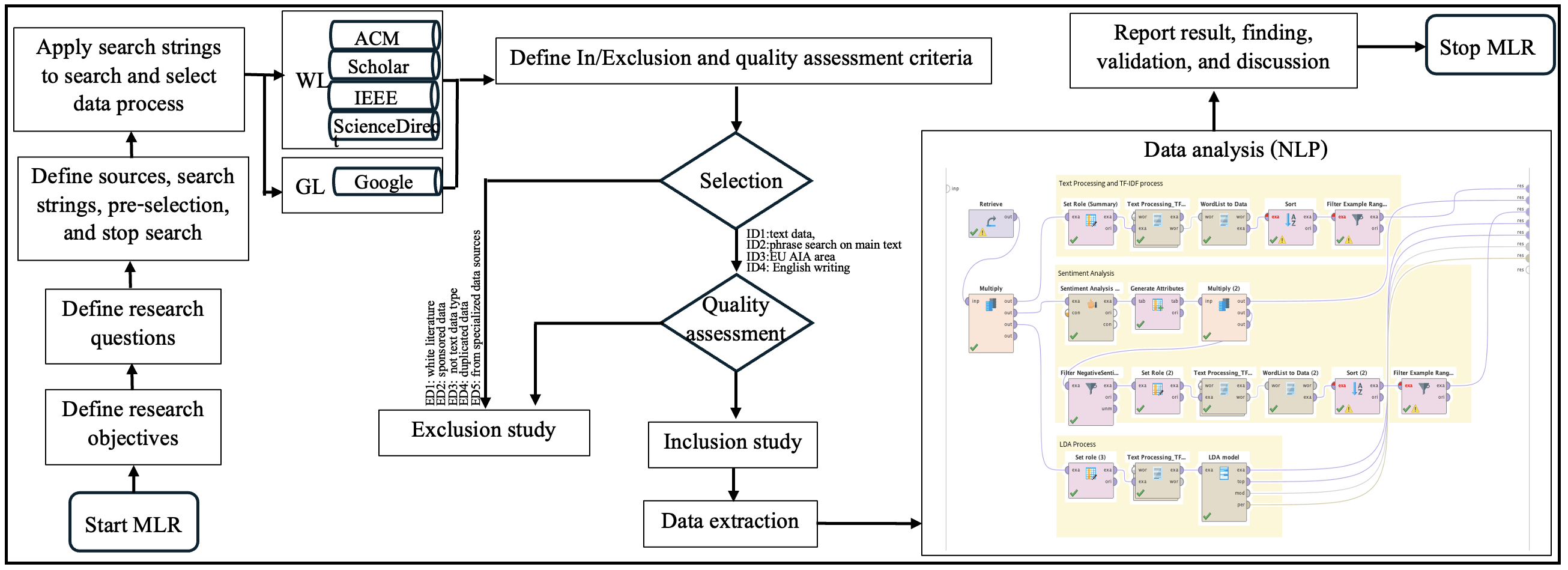}
	\caption{The system architecture of this study.}
	\label{fig:Process1}
\end{figure}

\emph{Text processing} implements several operations, such as: 1) token splitting to split text into individual words or tokens; 2) filtering stop-words to remove common words that do not carry significant meaning; 3) stemming to reduce words to their base root form; and 4) transforming cases to convert all text to lowercase.
\emph{Text vectorization processing} implements term frequency and the reciprocal document frequency (TF-IDF) operation to convert text into a numerical representation based on TF-IDF. 
\emph{Sentiment Analysis} is applied to determine the sentiment of the text, which can provide additional insights into the nature of the challenges.
\emph{Latent Dirichlet Allocation (LDA)} performs topic modeling and identifies underlying themes or topics in text data by applying an LDA model.
\emph{Visualization} utilizes 1) word clouds to visualize the most frequent terms or topics, and 2) graphs to represent results for better understanding.

\section{Results}
\label{result}
We report our results in this section, seeking to answer the RQs outlined in Section 1. The addition outlet types in this study are updated to the literature type of shades of GL with the same criteria.
The results of the selected and satisfied data sources are presented in subsection 3.1.
Lastly, the results of the analyzed data and validated outcome are presented in subsection 3.2.

\subsection{MLR execution, protocol and quality assessment}
\begin{figure}[ht]
	\centering
	\includegraphics[width=1.0\textwidth]{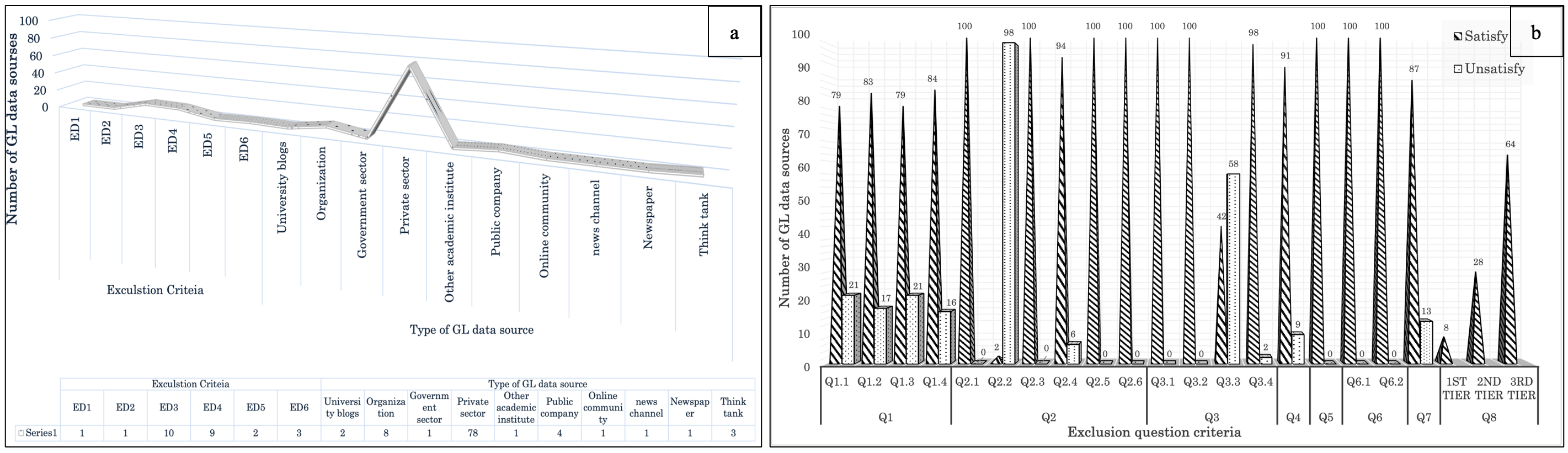}
	\caption{The results of a) the selected data sources from pre-selection and b) the results of data source quality assessment.}
	\label{fig:P1_2.png}
\end{figure}

Data was collected through a phrase search ("EU AI Act" AND "business"), pre-selection, and stopping criteria. The Google search yielded 214 million results, with relevant results across 30 pages. After pre-selection and applying stopping criteria, 100 results were found on the first 12 pages. Of these, 26 data sources were excluded based on various criteria and shown as Figure~\ref{fig:P1_2.png} (a).

The quality assessment criteria were applied to 100 data sources from the selected results. Based on these criteria, 26 sources were deemed suitable for inclusion as GL in this study. The excluded sources were primarily rejected due to vested interests or bias, lack of reputable authors or relevant expertise, failure to address specific questions, insufficient impact metrics, and outdated information. An example of the quality assessment is shown in Table~\ref{table:ResultQualityAssessment1} and Figure~\ref{fig:P1_2.png} (b).

\begin{table}[ht]
	\caption{Example of the quality assessment for GL data sources for EU AI Act in Business}
	\label{table:ResultQualityAssessment1}
  \scriptsize
  \begin{tabularx}{\textwidth}{lccccc X}
    \toprule
\multirow{2}{*}{EQ$^*$} & \multicolumn{5}{c}{GL Source} & \multirow{2}{*}{Note} \\
& A & B & C & D & E & \\
    \midrule
												
1.1 &1  & 1 &1  & 1 &  1 & All GL data sources are reputable publishing organizations \\
1.2 & 1  & 1 &1  & 1 & 1 & All GL data sources are published by an individual author\\
1.3 &1  & 1 &1  & 1 &  1 & All GL data sources have other work published in the field\\
1.4 &1  & 1 &1  & 1 &  1 & All GL data sources have expertise in the area\\ \midrule

2.1 & 1  & 1 & 1  & 1 &  1 & All GL data sources have a clearly stated aim\\
2.2  & 0 & 0 & 0 & 0 &0  & No GL data source has a stated methodology\\
2.3  & 1  & 1 & 1  & 1 &  1 & All GL data sources are supported by authoritative, contemporary references\\
2.4  & 1  & 0& 1  & 1 & 0 & All GL data sources, except B and E, have clearly statedlimits \\
2.5  & 1  & 1 & 1  & 1 &  1 & All GL data sources cover a specific question on EU AIA and business\\
2.6  & 1  & 1 & 1  & 1 &  1 & All GL data sources refer to a particular population or case in context\\ \midrule

3.1 &1  & 1 &1  & 1 &  1 & All GL data sources seem to be balanced in presentation\\
3.2 & 1  & 1 &1  & 1 &  1& The statements in all five GL data sources are objective and adhere strictly to factual information, avoiding subjective opinions\\
3.3 &  1  & 1 &1 & 1 &  0 & All GL data sources, except E, provide unbiased conclusions with no indication of vested interests. E promotes its compliance services, indicating a vested interest.\\
3.4 & 1  & 1 &1  & 1 &  1& The conclusions in all GL data sources are supported by the data\\\midrule

4.1 &1  &1 &1 & 0 & 0 & All GL data sources, expect D and E, have a clearly stated date \\ \midrule
  	
5.1 &1  &1 &1 & 1 & 1 & All GL data sources discuss the EU AI Act thoroughly, referencing specific provisions, expert opinions, and impacts on the tech industry. They include related sources and formal documents, examining implications for businesses, compliance requirements, and regulatory alignment.\\\midrule
6.1 & 1  & 1 &1  & 1 &  1& All GL data sources provide detailed guidance for businesses on EU AIA compliance, with sector-specific insights, particularly for financial services and UK businesses. They offer practical steps for implementing human oversight and improving business competitiveness\\
6.2 & 1  & 1 &1  & 1 &  1&All GL data sources emphasize the need for comprehensive AI regulations for ethical deployment and proactive compliance strategies for the EU AIA. They highlight balanced regulation to avoid excessive burdens, international alignment, and human oversight to ensure safety and compliance.\\ \midrule
7.1 &1  & 1 &1  & 0 &  1& All GL data sources, except D, provide impact metrics. Two impact metrics are used for analysis, namely number of backlinks  and social media shares.\\  \midrule
8.1 & .5 & 0 & 1 &0 &0 & GL data source A is an \emph{Informational Article}, B is an \emph{Informational Blog Post}, C is a \emph{Position Paper}, D is an \emph{Advisory Article}, and E is a \emph{Blog Post}\\ \midrule
Nor.			& .9 &.8 & .9  & .7  & .8  & Normalized mean value of the sum of scores, rounded up\\
    \bottomrule
    \multicolumn{7}{l}{$^*$ see criteria and EQ explanations from Table~\ref{table:QualityAssessmentCriteria}}
    \end{tabularx}
\end{table}

\clearpage
Regarding the quality assessment of authority, all Exclusion Questions (EQs) related to authority were satisfied. In the methodology quality assessment, twenty EQs were examined. All data sources qualified as GL based on five of these questions. However, question 2.2, regarding stated methodologies, was omitted since formal methodologies are not typically used in these articles. Regarding objectivity, 40 sources passed the quality assessment. Specifically: most of the texts provided a balanced view, discussing both challenges, benefits, statements were generally objective, focusing on facts and expert opinions, and conclusions free of bias / vested interest in terms of the EU AIA (EQs 3.1--3.4). Concerning date quality assessment, 91 articles have clearly stated publication dates, ranging from August 2021 to May 2024, as per EQ 4.1.
With respect to related sources quality assessment, all articles discuss and are linked to key regulatory frameworks, including the EU AIA, as assessed by EQ 5.1.
In the novelty quality assessment, each article offers unique insights or practical advice on EU AIA compliance, enhancing current understanding with detailed explanations and updates, as assessed by EQs 6.1 and 6.2.
In the impact quality assessment, EQ 7.1, each article's impact was normalized into a single aggregated metric. Professional networks were assessed via LinkedIn, Wikipedia, and Trustpilot. Backlink counts were determined using Ahrefs, and social media shares were measured using SharedCount.

Some finding outcome types in this study are not listed in the "shades of GL". These were classified by shades of GL criteria by using expertise and outcome control, equivalent to the criteria used. The additional classification outcome types are shown in Table~\ref{table:AdditionalClassification}.

\begin{table}[ht]
    \caption{The additional classification type of GL data sources based on Shades of GL}
    \centering
    \scriptsize
    \begin{tabularx}{\textwidth}{lp{4.5cm}X}
    \toprule
    GL Tier & Shades of GL \cite{garousi2019guidelines}  & Additional classification types in this research study\\
   \midrule
   $ 1^{st} $ tier & Book, full paper, position paper, abstract, poster, conference paper & Expert commentary, position paper, analysis article, insights legal analysis and guidance article\\
   $ 2^{nd} $ tier & Annual reports, news articles, video, Q/A site (e.g. StackOverflow), Wiki article & Client alert publication, insight article published, information report, legal advisory publication, advisory publication\\
   $ 3^{rd} $ tier & Blog posts, emails, tweets, presentations	& Information article, information blog post, legal advisory blog post, consulting report, opinion blog post, business insights article\\
    \bottomrule
    \end{tabularx}
    \label {table:AdditionalClassification}
\end{table}

\subsection{Interpreted insights from the Data Analysis Approach}
\begin{figure}[ht]
	\centering
	\includegraphics[width=1.0\textwidth]{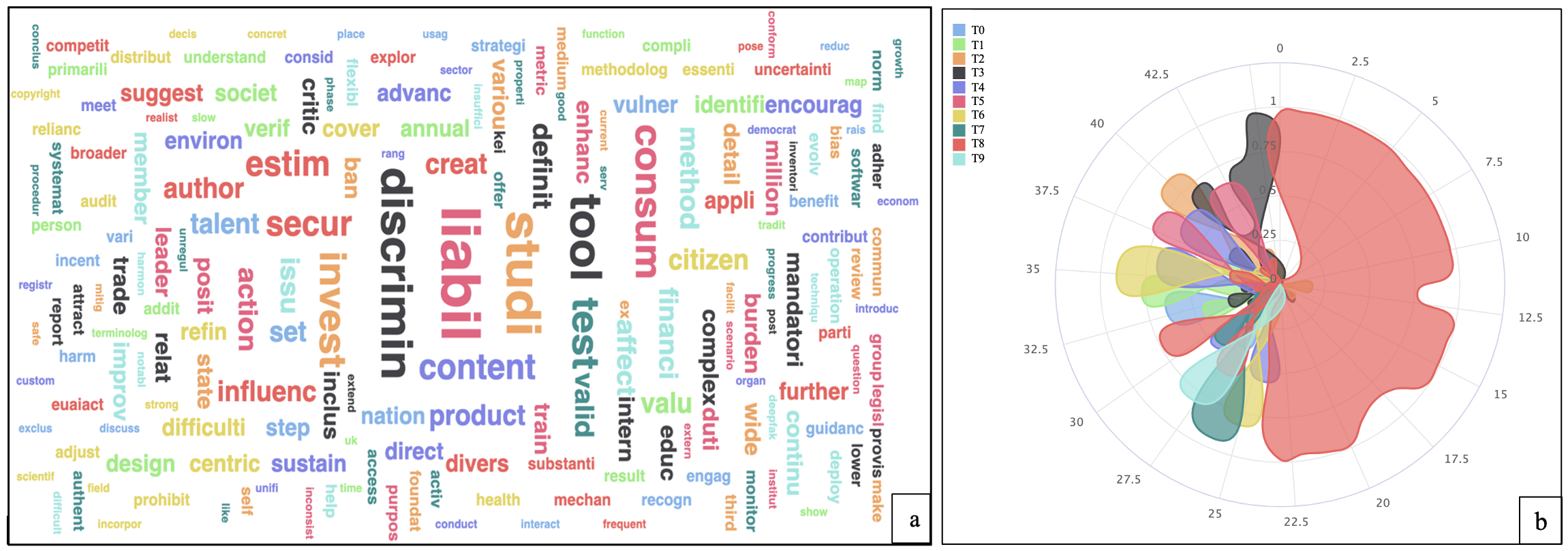}
	\caption{The extracted results of a) the challenges and b) the strategies and processes required in complying with the EU AIA.}
 	\label{fig:img1}
\end{figure}

\emph{RQ1: What are the key challenges perceived by both industry and academia in complying with the EU AIA?}
In this approach, NLP, a text processing TF-IDF method, was implemented. The analysis was conducted on the text dataset, analyzing 199 words. This method was used to identify the most significant terms in the documents by considering both their frequency within individual documents (Term Frequency) and their rarity across all documents (Inverse Document Frequency). Through this approach, three key challenges were identified, i.e., liability, discrimination, and the adequacy of the tool.
"Liability" with the highest TF-IDF scores was considered the most challenging and relevant to specific documents (total occurrences: document frequency: 10:2). The insight finding revealed that enterprises need to adapt their compliance and liability management. The second and third challenges are discrimination and tools, i.e., enterprises must manage discrimination risk both for legal compliance and to maintain customer trust and the adequacy of the tool. The 199 extracted primary keywords are visualized  in Figure~\ref{fig:img1} along with their frequency for both industry and academia when complying with the EU AIA.

Furthermore, the results from sentiment analysis identified a few negative points of concern when enterprises, both in industry and academia, comply with the EU AIA regulation, in terms of requirements and transparency in the AI system, mainly in the high risk level category. Concerning regulation, the findings suggest implementing a risk-based classification with stringent requirements, with the aim of conducting thorough assessments and compliance checks as laid down in the Act's regulatory framework. The findings identified requirements such as: 1) balancing innovation with safety and ethical standards through specific compliance measures, and 2) updating AI strategies and categorizing AI solutions to ensure compliance with the new regulations. In addition, concerns were raised about transparency. The insights suggest that a human rights-based approach should be emphasized in risk management and human oversight [GL10, GL17, GL21].

\emph{RQ2: What strategies and processes are enterprises developing to implement the EU AIA?}
To address this question, NLP, i.e., the text processing TF-IDF method and LDA, was implemented for analyzing the underlying themes of strategies and processes being developed to implement the EU AIA. Through this approach, the model analyzes ten themes regarding strategies and processes, with each theme represented by a set of words most strongly associated with it. The analyzed results also revealed that our contexts cover more topics evenly and have more commonality between topics, which are indexed by the alphaSum (0.9008) and beta (0.1582) indicators.

Figure~\ref{fig:img1} represents the 10 themes of strategies and processes being developed to implement the EU AIA, where the wider areas of the themes are addressed by the tokens index. Then, the coherence index was applied to underline three crucial themes: 1) \emph{T8: risk-based regulatory compliance systems}, 2) \emph{T3: ethical frameworks and principles in technology development}, and 3) \emph{T4: policies and systems for regulatory risk management}. Each theme is described below:

\emph{T0: Proposed legislation and financial liabilities} seems to focus on themes related to liability, financial implications, directives, and legislation. The presence of words like "liability," "financial," and "legislation" with high exclusivity scores suggests that these are central to the theme of this topic.

\emph{T1: Regulatory standards and testing processes for consumer protection} appears to revolve around themes of standards, testing processes, and regulatory concerns. Despite the strong coherence score, the high exclusivity for words like "test" and "vulner" suggests that these are key aspects of the topic.

\emph{T2: Studying discrimination and trust in regulatory demand} focuses on themes related to discrimination, demand, and trust. The high exclusivity of "discrimin" and "rbi" suggests a specific focus on issues surrounding discrimination and a regulatory body or measure.

\emph{T3: Ethical frameworks and principles in technology development} centers on ethics, practices, and frameworks, particularly in the context of development. The high exclusivity of terms like "ethic" and "practic" indicates a strong focus on ethical principles and best practices within this topic.

\emph{T4: Policies and systems for regulatory risk management} revolves around regulatory frameworks, policies, and risks. The terms "regulatori" and "act" are central to the theme, while the lower exclusivity of "system" and "risk" indicates that these concepts are prevalent across multiple topics. 

\emph{T5: User information and labeling systems impact} focuses on information systems, impacts, and labeling. The high exclusivity of "inform" and "label" suggests these are central concepts to the topic, while "system" appears to be a common term across multiple topics. 

\emph{T6: Cost estimation and compliance in system development} related to costs, compliance, and estimates. The high exclusivity of "cost" and "estim" indicates a strong focus on financial or resource-related aspects within this topic.

\emph{T7: Global market dynamics and investment in European talent} focuses on themes around markets, investments, and talent. The terms "invest" and "talent" have high exclusivity, suggesting a strong emphasis on investment strategies and human resources, particularly in a global or European context. 

\emph{T8: Risk-based regulatory compliance systems} is centered on risk, compliance, and regulatory requirements. The term "risk" has high exclusivity, indicating that risk management and assessment are primary concerns in this topic.

\emph{T9: Justice and bias in sustainable development goals} is highly concentrated on Sustainable Development Goals (SDG), justice, and related terms. The high exclusivity of "sdg" and "justic" indicates that this topic is very specialized, likely addressing a niche but important area within the broader context.      

To validate these outcomes, we crosschecked our results with other existing academic publications. A similar trend of key challenges was found in Arcila, B.B., Gerke S. et al., and Van K.H. [WL18]\cite{gerke2020ethical,van2022eu} as there is an effect on liability when an AI system is harmful, which then affects and harms fundamental rights and safety. Then discrimination, privacy, and decision-making are impacted after fundamental rights. The publication by Foffano et al. highlights: 1) ethical principles and policies as a part of the main AI strategy for the social good based on the perspective of the EU; 2) compliance management, risk-based compliance systems, and regulation for enterprises to lead their organizations\cite{foffano2023investing,savin2023strategic}.
 
\section{Discussion}
\label{discussion}

\begin{figure}[ht]
	\centering
	\includegraphics[width=1.0\textwidth]{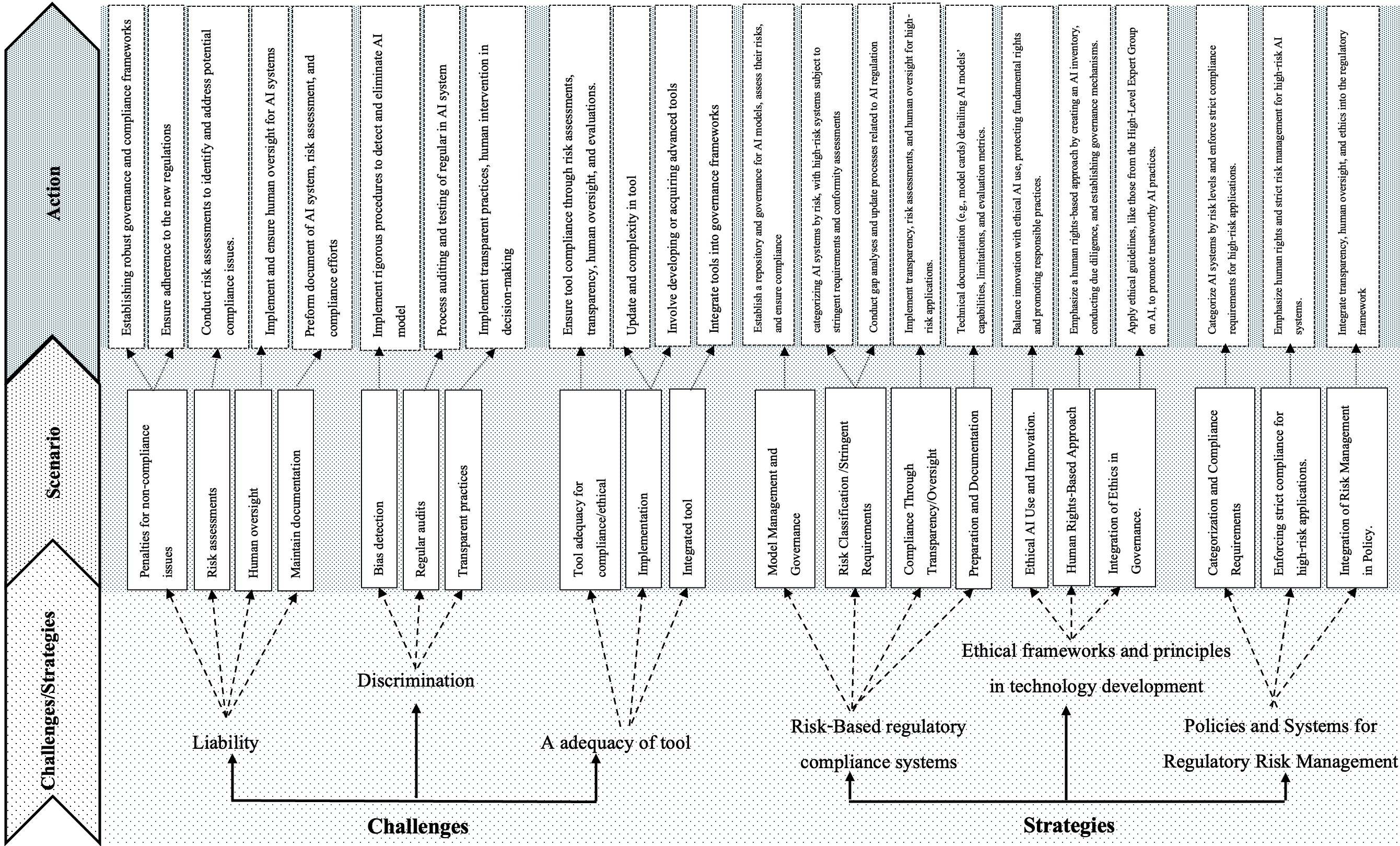}
	\caption{The framework outline of key challenges and strategies when complying with the EU AIA.}
	\label{fig:FW}
\end{figure}

This section comprises a detailed discussion of the extracted insights on: 1) the three key challenges---liability, discrimination, and tools, and 2) the three crucial themes for developing strategies to implement the EU AIA---risk-based regulatory compliance systems, ethical frameworks in technology, and regulatory risk management systems. We then outline the framework of key challenges and strategies for complying with the EU AIA, as shown in Figure~\ref{fig:FW}. Lastly, we discuss the limitations of this study. We begin by discussing the three key challenges:

\subsection{Key challenges}

\emph{Liability}. Insight extraction revealed that enterprises need to adapt their compliance and liability management. Based on this finding we divided how to deal with it into the following four key areas: 1) Penalties for non-compliance issues. This issue creates significant financial liability for businesses that fail to comply with regulations. To address this, enterprises must: i) proactively establish governance and compliance frameworks to mitigate risks, and ii) ensure adherence to the new regulations; 2) Risk assessments. Thorough risk assessments for identifying and addressing potential compliance issues should be conducted [GL2];
3) Maintain documentation. By documenting AI systems, risk assessments, and compliance efforts, the records serve as evidence of compliance and help mitigate liability during regulatory scrutiny; and 4) Human oversight. To mitigate liability risk, human oversight mechanisms for AI systems need to be implemented and ensured, particularly in critical decision-making areas like healthcare, where errors can have severe consequences [GL8].

\emph{Discrimination}. 
Managing discrimination risk is important for both legal compliance and maintaining customer trust. If an enterprise fails regarding discrimination, it could result in penalties under the EU AIA and also damage the company's reputation and trust from customers. Therefore addressing discrimination proactively is crucial for compliance and maintaining public trust. Insight extraction revealed three key parts for managing discrimination risk, as follows: 1) Bias detection. Perceptions of discrimination strongly influence the demand for regulation. Implementing rigorous procedures to detect and eliminate biases in the AI model will minimize discrimination. This includes using diverse and representative datasets during the training phase to prevent discriminatory outcomes [WL22]; 2) Regular audits. The process of regular auditing and testing in AI systems is important and recommended as it leads to identification of any unintended discriminatory patterns that may emerge over time [GL34]; and 3) Transparent practices. Such practices are crucial to represent where the accountability for any discriminatory outcomes lies. To implement transparent practices, it could be a benefit to include human intervention in the decision-making process to ensure fairness \cite{Sophie2023}. Furthermore, one extracted insight was the integration of non-discrimination and fairness into AI development practices, as these are beneficial to fill the gaps between AI ethical guidelines and industry practices [WL29]. 

\emph{Adequacy of tool} insight indicates that effective, adaptable, and compliant tools are required to address the challenges faced when complying with the EU AIA, as well as ensuring that enterprises can meet the regulatory demands and maintain ethical AI practices. The insight extraction of three key parts concerning tool adequacy is as follows: 1) Tool adequacy for compliance and supporting ethical AI practices. To ensure that they have the appropriate tools to support ethical AI practices, this insight proposes enterprises use risk assessments, transparency, human oversight, other assessments, and compliance checks [GL21, GL54]; 2) Implementation, updating and complexity of the tool. This insight highlights the need for adaptation, updates, and tool complexity to meet evolving compliance requirements, especially as the EU AIA introduces new regulatory frameworks, and to manage the complexities of AI systems. Enterprises should ensure their tools undergo risk assessments, transparency evaluations, human oversight, and other compliance checks [GL10, GL18]; and 3) An integrated tool for governance and risk management. This insight underlines the benefits of effective AI system management and helps enterprises meet EU AIA compliance standards. Enterprises should integrate tools into governance by establishing  frameworks to mitigate risks and ensure adherence to the new regulations [GL2].

\subsection{Themes for strategy development}

Next, we discuss the three crucial themes for developing strategies and processes to implement the EU AIA:

First theme: \emph{Risk-based regulatory compliance systems}. This theme highlights the central focus on risk followed by compliance, system, and regulatory requirements. It suggests that this theme is concerned with systems that are designed to ensure compliance with regulatory requirements through a risk-based approach. The analysis results pointed to five ways to help enterprises ensure compliance with regulatory requirements through a risk-based approach such as 1) Model Management and Governance. The key point is that enterprises must establish a repository and governance for AI models, assessing their risk levels and ensuring compliance [GL2]; 2) Risk Classification and Stringent Requirements; 3) Focus in AI systems is categorized by risk, with high-risk systems subject to stringent requirements and conformity assessments. This suggests that enterprises must conduct gap analyses and update processes related to AI regulation [GL34, GL38, GL49]; 4) Compliance through Transparency and Oversight. This refers to implementing compliance requirements in terms of transparency, risk assessments, and human oversight for high-risk applications [GL29, GL32].
In the case of SMEs, they may face significant costs and legal uncertainty (compliance burden) [GL41]; and 5) Preparation and Documentation. This recommends that enterprises integrate ethical charters, legal tools, and technical documentation to establish a comprehensive governance framework for AI. Technical documentation, such as model cards, is vital for auditing AI systems, ensuring transparency and educating users on proper usage. It details AI models' capabilities, limitations, and evaluation metrics. For example, the Big Science Workshop illustrated how collaborative AI development can effectively incorporate ethical, legal, and technical compliance [WL24].

Second theme: \emph{Ethical frameworks and principles in technology development}. This theme underscores the importance of ethical frameworks and principles in technology development, highlighting how ethical considerations guide the practical creation of technological solutions. The analysis identified three approaches to assist enterprises in addressing ethical frameworks and principles in technology development, as follows: 1) Ethical AI Use and Innovation. This proposes that enterprises implement ethical AI use while balancing innovation which i) protects fundamental rights, ii) promotes responsible AI practices, iii) focuses on ethics, safety, transparency, and human oversight to protect democratic processes [GL34, GL37, GL38]; 2) Human Rights-Based Approach. This emphasizes that a human rights-based approach by enterprises should be prepared by creating an AI inventory, conducting human rights due diligence, and establishing governance mechanisms [GL17]; and 3) Integration of Ethics in Governance. In this theme, enterprises are advised to apply ethical guidelines from a high-level expert group on AI integrated into the regulation, promoting trustworthy AI practices [WL9].

Third theme: \emph{Policies and systems for regulatory risk management}. This theme highlights the central topic of regulatory acts and systems, and the policy put in place for managing risks. It suggests a focus on how regulations guide the development and implementation of risk management systems within various policy frameworks such as 1) Categorization and Compliance Requirements. This outlines how to categorize AI systems by risk levels and imposes strict compliance requirements, especially for high-risk applications [GL23], 2) Enforcing strict compliance for high-risk applications, 3) Emphasis on Human Rights and Risk Management by implementing strict requirements for high-risk AI systems, with the emphasis on compliance in term of transparency, risk management, and human oversight [GL39, GL49]; and 4) Integration of Risk Management in Policy. This encourages enterprises to integrate the transparency, human oversight, and ethical considerations in the regulatory framework [WL21].

\textbf{Limitations.} The empirical results reported herein should be considered in the light of some limitations. We divided these limitations into two types, i.e., internal and external threats. The internal threat of this study is related to 1) how we classified and interpreted the type of GL and 2) how we set the normalization impact metrics of GL in quality assessment for the literature review approach. For example, some GL types were not indicated in the GL classification, and some of the information was not always easy to retrieve. To mitigate this threat, we applied "shades of gray" criteria to classify the type of GL\cite{garousi2019guidelines}, we manually counted the number of citations, backlinks, social media shares, and the number of comments posted, and we manually checked the author's reputation and/or expertise in the area. This problem was also found in GL\cite{kamei2021grey}. A minor internal threat, reducing words to their root forms in text processing can obscure differences between word forms. Mitigate this by manually rechecking the original words or sentences. One external threat of this study is related to the methodology, objectives, and impact criteria in quality assessment for the literature review approach. For example, not all of the GL data was stated in formal methodology, which is typically used in this literature, and although some of the GL data was not presented objectively, it had an impact index (n=74). Therefore, one criterion in RQ 2.2 was omitted and GL without an objective and impact index was not used in the data analysis approach. Concerning the omitted methodology criteria, it was necessary to use those samples as the GL in question contained the most relevant data. However, comparing the sample, including threats from the objective and impact index, we checked the context manually and showed quite a similar trend for the interpreted results.

\section{Conclusion and Future Work}
\label{conclusionandfuturework}

We conducted an MLR and utilized NLP to help analyze the academic articles and gray literature on the EU AI Act in business contexts to aid enterprises in complying with and benefiting from the EU AIA. We examined 56 of 130 articles for comprehensive understanding, identifying 199 keywords, including liability, discrimination, and tool adequacy, as major obstacles for enterprises. We also noted negative aspects in the articles, such as regulatory interpretations, specific requirements, and transparency issues of the Act. Additionally, we identified ten strategic themes, with three key areas: risk-based regulatory compliance, ethical frameworks in technology development, and policies for regulatory risk management. We validated our findings against existing academic literature and discussed the results while establishing a practical framework for EU AIA compliance. These findings will benefit enterprises and research in the EU AIA domain. In future work, the focus is more likely to be on sectors with specific challenges and strategies such as "Deep Dive into High-Risk Sectors," with the implementation of AI in healthcare, finance, or transportation. This could involve exploring how these sectors are preparing for compliance and the unique challenges they face.

\subsubsection*{Declaration of competing interest}
The authors have no competing interests to declare that are relevant to the content of this article.
\subsubsection*{Data availability}
A list containing all the references of this study is available at \url{https://doi.org/10.6084/m9.figshare.26968498.v2}.

%
%



\bibliographystyle{plain}
\bibliography{main.bib}

\end{document}